\documentstyle[11pt,newpasp,twoside,epsf]{article}
\begin{document}
\title{Evolution of an afterglow with a hard electron spectrum}
\author{D Bhattacharya}
\affil{Raman Research Institute,Bangalore - 560 080, India}
\author{L Resmi}
\affil{Raman Research Institute,Bangalore - 560 080, India}

\begin{abstract}
Diffusive shock acceleration theory suggests that the ``universal'' 
energy spectrum of electrons with a power-law index $p\approx 2.3$,
commonly used to model GRB afterglows, cannot extend below an electron 
lorentz factor $\gamma_e$ equal to the bulk lorentz factor of the blast 
wave multiplied by the ratio of proton and electron masses.  We
suggest that the electron energy distribution has a slope $p<2$ 
below this limit, down to an appropriate $\gamma_m$.  A two-slope
spectrum such as this provides a good model for the afterglow
of GRB010222.
\end{abstract}

\section{Introduction}
GRB afterglow emission is usually modelled with a single
power-law electron energy spectrum:
\begin{equation}
N(\gamma_e) \propto \gamma_e^{-p}\;\;\;\gamma_m<\gamma_e<\infty 
\end{equation}
with the power-law index $p > 2$ and $\gamma_e$ being the lorentz
factor of the electron (see Piran 1999 for a review).  Relativistic 
shock acceleration theory predicts an universal $p \approx 2.3$ which 
fits many afterglow spectra (Waxman 1997; Galama et al 1998; Wijers and 
Galama 1999).

In some GRB afterglows, harder electron energy spectrum $p < 2$
appears to be needed (Panaitescu and Kumar 2001a; Sagar et al 2001; Cowsik 
et al 2001; Bhattacharya 2001).  Such a spectrum must have an upper 
cutoff, an {\em injection break $\gamma_i$}, beyond which $p$ 
must exceed $2$ in order to keep the total energy in the electron 
distribution finite.  The evolution of the afterglow emission is
influenced by the evolution of $\gamma_i$ (Bhattacharya 2001; Panaitescu
and Kumar 2001b; Dai and Cheng 2001)

The minimum lorentz factor $\gamma_m$ of the distribution is
determined by the fraction $\epsilon_e$ of the postshock thermal
energy that is injected into electrons.  For $p > 2$ this is
given by
\begin{equation}
\gamma_m = \epsilon_e\left(\frac{p-2}{p-1}\right)\cdot
              \left(\frac{m_p}{m_e}\Gamma_{\rm sh}\right) 
\end{equation}
where $\Gamma_{\rm sh}$ is the bulk lorentz factor of the blast
wave shock and $m_p$, $m_e$ are proton and electron masses
respectively (Sari, Piran and Narayan 1998).

\section{An injection break}
Relativistic shock acceleration theory predicts that the
{\em universal spectrum} with $p \approx 2.3$ can extend only
down to a minimum $\gamma_{\rm acc} = (m_p/m_e)\Gamma_{\rm sh}$.
Below this energy the electron Larmor radius becomes smaller 
than the shock thickness, the shock is no longer perceived as
a discontinuity and the acceleration efficiency drops
(Hoshino et al 1992; Gallant et al 2000; Kirk 2002).  
However for a low $\epsilon_e$ the $\gamma_m$ estimated from 
eq.~2, and routinely used in GRB afterglow modelling,
works out to be much lower than $\gamma_{\rm acc}$.
This creates a contradiction.

In the region between $\gamma_m$ and $\gamma_{\rm acc}$, some special 
pre-acceleration mechanisms are needed (Kirk 2002; Gallant et al 2000).
One candidate for this is a variant of the ion magnetosonic wave absorption
mechanism proposed for electron-positron-proton plasma (Hoshino et al
1992). This mechanism produces a hard energy spectrum, but has a 
natural upper cutoff at exactly the $\gamma_{\rm acc}$ defined above.
We conjecture that for electrons in GRB afterglows, a $p_2 > 2$
universal spectrum is valid
down to $\gamma_{\rm acc}$, and the energy spectrum below this,
down to an appropriate $\gamma_m$, is harder, with a $p_1 < 2$.
We identify the injection break $\gamma_i$ with $\gamma_{\rm acc}$.
This would also introduce a corresponding injection break $\nu_i$
in the radiation spectrum. Here
\begin{equation}
\gamma_m=\left[ \epsilon_e \frac{(2-p_1)(p_2-2)}{(p_2-p_1)(p_1-1)}
         \right]^{1/(p_1-1)} \gamma_i
\end{equation}

\section{Evolution of the radiation spectrum}
We have calculated the evolution of the afterglow radiation spectrum
based on this premise.  Writing the flux at a frequency $\nu$ at time $t$
as $F_{\nu}(t) \propto t^{\alpha}\nu^{\beta}$, we find the values of
$\alpha$ and $\beta$ in different spectral regimes to be as follows:

\begin{table}
\caption{light curve and spectral slopes of an afterglow with a
         two-slope electron energy distribution: $p_1 <2$ and 
         $p_2 >2$}
\begin{center}
\begin{tabular}{|cccc|}\tableline
                  & $\beta$ & $\alpha_1$ & $\alpha_2$ \\ \tableline
$\nu<\nu_a$       & $  2  $ & $ 1/2  $   & $ 0 $      \\ 
$\nu_a<\nu<\nu_m$ & $ 1/3 $ & $ 1/2  $   & $ -1/3 $   \\ 
$\nu_m<\nu<\nu_c$ &$-(p_1-1)/2$&$-3(p_1-1)/4$ & $-p_1$ \\ 
$\nu_c<\nu<\nu_i$ &$-p_1/2$  &$-(3p_1-2)/4$&$-p_1$     \\ 
$\nu_i<\nu$       &$-(p_2/2)$ &$-(3p_2-2)/4$&$-p_2$    \\ 
\tableline\tableline
\end{tabular}
\end{center}
\end{table}

In the table, the third column gives the light curve slopes after
the lateral spreading of the collimated outflow dominates the
dynamics, and the second column those before this phase.  The
transition between them is often called the {\em jet break}. The
quantity $\nu_c$ stands for the synchrotron cooling break frequency
in the radiation spectrum.

\begin{figure}[h]
\plottwo{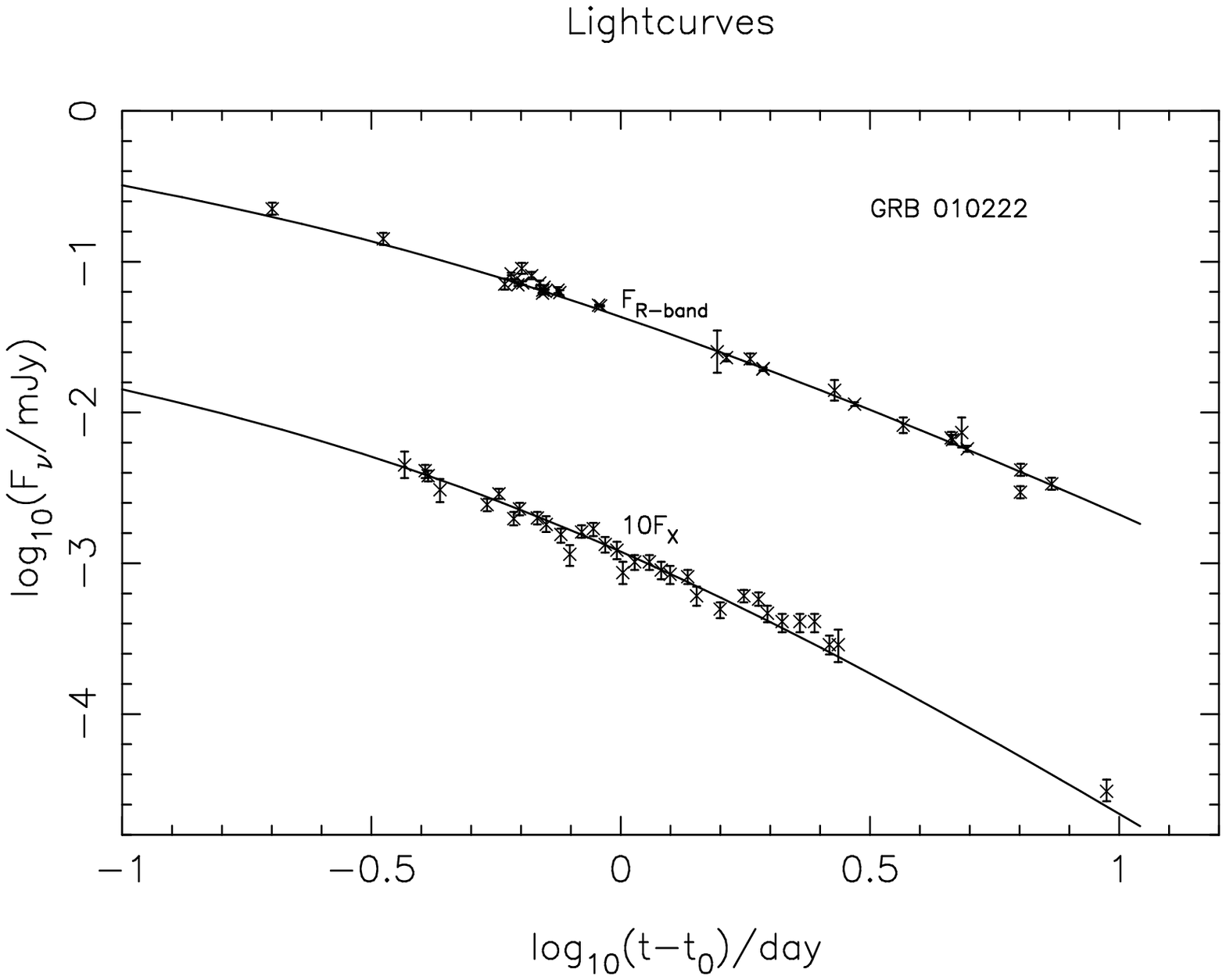}{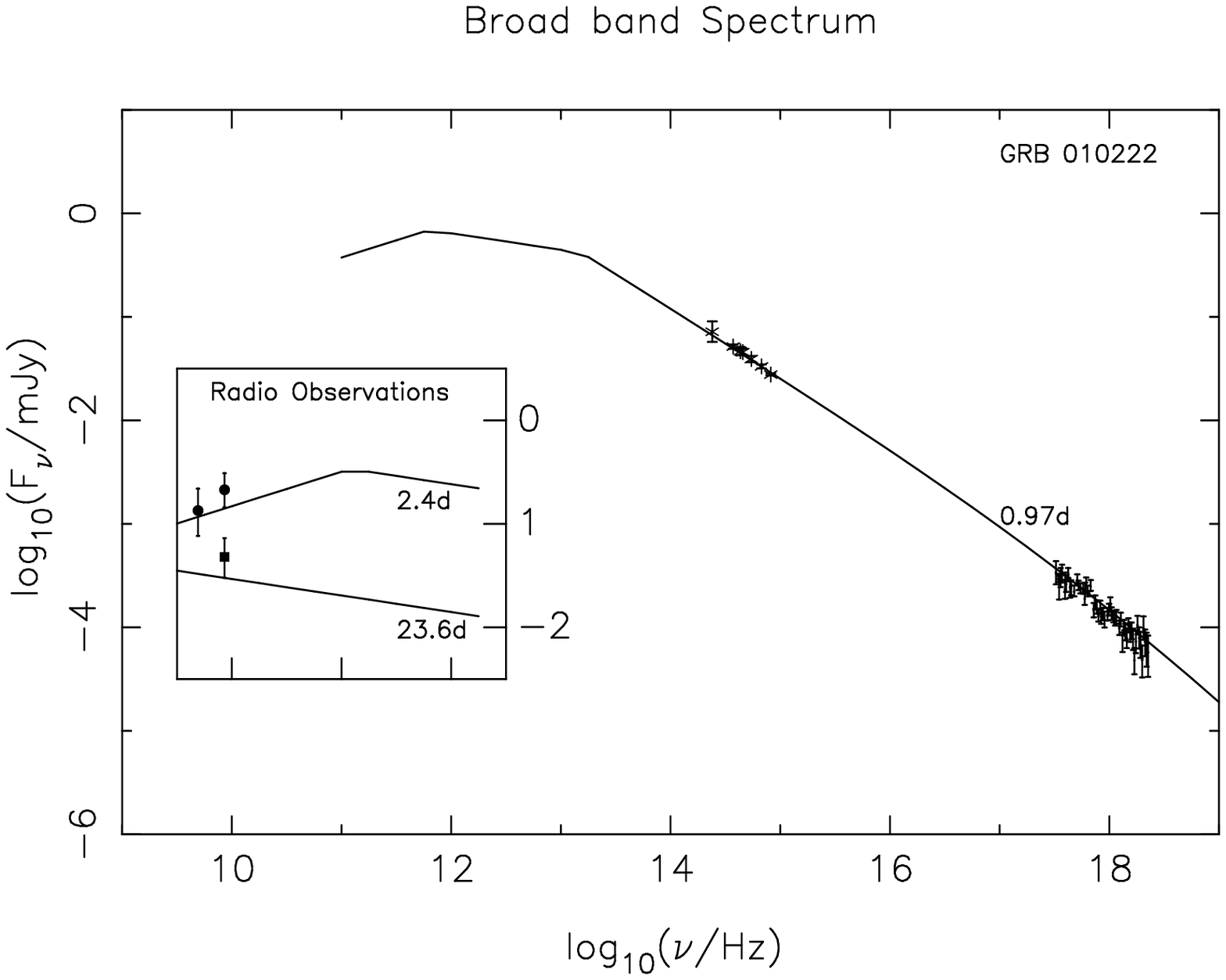}
\caption{Light curve and spectra of the GRB010222 afterglow 
compared with the model (solid lines) using a two-slope electron energy 
spectrum.  Optical data taken from the compilation of Sagar et al
(2001) and Cowsik et al (2001), X-ray data from in't Zand et
al (2001) and Radio data from Frail et al (2002).  The model uses  
$p_1=1.32$, $p_2=2.1$ and a jet break at 0.4 days. The injection break
$\nu_i$ is located at $\sim 10^{18}$~Hz (X-rays) $\sim 1$~day after the 
burst.}
\end{figure}

\section{GRB010222}
The afterglow of GRB010222 has been modelled previously
as a burst in a very high density medium undergoing a transition to a
non-relativistic expansion (Masetti et al 2001), which cannot explain 
the early appearance of radio emission; or as a hard electron spectrum 
afterglow that underwent an early jet break (Sagar et al 2001; Cowsik et 
al 2001; Panaitescu and Kumar 2001a), which had difficulty explaining the
spectral slope observed in X-ray bands.

We suggest that this afterglow had an injection break $\nu_i$ in the
X-ray band $\sim 1$~day after the burst, had a jet break transition
at $t_j\sim 0.5$~day, and evolved in a normal ISM. With an assumed
$p_1=1.3$, $p_2=2.1$, and a smooth joining of the power-law segments,
we are able to obtain good fits to the spectrum and the light curve
of this afterglow, as shown in the figures.

\section{Summary}
We have computed the evolution of a GRB afterglow with a hard electron
energy spectrum up to an injection break at $m_p/m_e$ times the bulk
lorentz factor of the shock.  Above the injection break, relativistic
shock acceleration predicts an universal, relatively softer, electron
energy spectrum. We obtain good fits to the observed evolution of
GRB010222 afterglow using this model.  We suggest that hard spectrum
emission is more likely to be seen in GRB afterglows that have a relatively
low $\epsilon_e$.

\end{document}